\documentclass[preprint2]{aastex}
\usepackage{natbib}
\shorttitle{Particle acceleration in  4C74.26}
\shortauthors{Araudo et al.}

\begin{document}

\title{Particle acceleration and magnetic field amplification
in the jets of 4C74.26}

\author{A.T. Araudo\altaffilmark{1}, A.R. Bell\altaffilmark{2} and 
K.M. Blundell\altaffilmark{1}}

\altaffiltext{1}{University of Oxford, Astrophysics, Keble Road, 
Oxford OX1 3RH, UK}

\altaffiltext{2}{University of Oxford, Clarendon Laboratory, Parks Road, 
Oxford OX1 3PU, UK}

\begin{abstract}
We model the multi-wavelength emission in the southern hotspot of the  
radio quasar 4C74.26. The synchrotron radio emission is resolved near
the shock with the MERLIN radio-interferometer, and the rapid decay of  
this emission behind the shock is interpreted as the decay of the 
amplified
downstream magnetic field as expected for small scale turbulence.
Electrons are accelerated to only $0.3$~TeV, consistent with a diffusion
coefficient many orders of magnitude greater than in the Bohm regime.
If the same diffusion coefficient applies to the protons, their maximum energy
is only $\sim 100$~TeV.
\end{abstract}


\section{Introduction}

Diffusive shock acceleration (DSA) is an established mechanism to
convert bulk kinetic energy into a non-thermal distribution of
relativistic particles with a maximum energy much larger than the
average energy of particles in the plasma. This theory explains well
the spectrum of Galactic cosmic rays (CR)  with energies up to
$\sim$3~PeV, accelerated in supernova remnant shocks 
\citep[see][for a review]{Bell_rev_14}.  The most energetic CR,
i.e. particles with energies up to $50$~EeV, are accelerated outside
the Galaxy but the origin of these particles is still
unknown. Relativistic shocks in extragalactic sources like Gamma Ray
Bursts  and Active Galactic Nuclei  have been proposed as candidates
\cite[e.g.][]{Gallant,Murase}.  However, theoretical models
\citep{Pelletier, lemoine-pelletier-10, Sironi_13, Brian_14} show 
magnetic field amplification at  ultra-relativistic shocks on
scales much smaller than the Larmor radius $r_{\rm g}$ of particles
being accelerated  which precludes CR acceleration to EeV energies unless
other processes can be found to amplify the magnetic field on larger scales. 

Hotspots are usually detected at the jet termination region in
type II Fanaroff-Riley (FR) radiogalaxies \citep{FR}. The location
of the hotspot is coincident with the downstream region of the jet
termination shock, where particles accelerated by the shock emit
synchrotron radiation. Therefore, hotspots are suitable places to
study DSA in high velocity shocks.

We model the emission from radio to X-rays in the southern hotspot of
the FR~II source 4C74.26  using  data provided in
\cite{Erlund_TwoShocks}. We determine that the compact radio
emission traces out the location of the shock where the magnetic
field $B$ is amplified by plasma instabilities up to
$\sim$100~$\mu$G, and it damps rapidly
downstream of the shock.  The turnover in the synchrotron spectrum
between infrared  (IR) and optical wavelengths implies that the CR
electron scattering length is much longer than the Larmor radius
and consistent with the amplified magnetic field being structured on
very small scales and comparable with the ion skin-depth 
$c/\omega_{\rm pi}$.  
Cosmic ray acceleration is consequently very slow and electrons are 
accelerated to only $E_{e,\rm max} \sim 0.3$~TeV.  A similarly low acceleration
rate for ions would limit their energy to $\sim 100$~TeV
and preclude proton acceleration to EeV energies.

\section{The giant FR II galaxy 4C74.26}

\begin{figure}
\vspace{-5cm}
\includegraphics[width=0.4\textwidth]{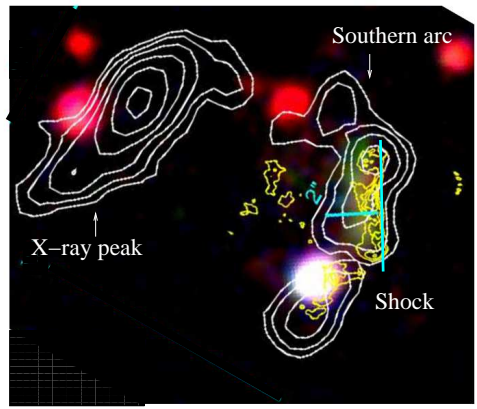}\\
\includegraphics[width=0.4\textwidth]{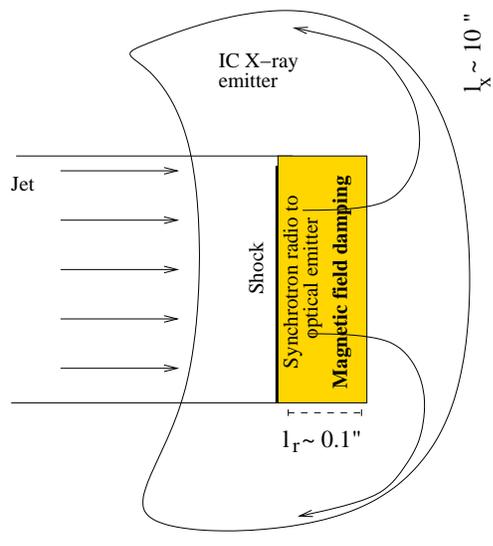}
\caption{\emph{Upper:} the hotspot(s) at the southern jet of
the FR~II galaxy 4C74.26 \cite[adapted and rotated by
$\sim$60$^{\circ}$ from][]{Erlund_TwoShocks}. 
White and yellow contours are X-rays and radio data, respectively.
Red and green correspond to IR and optical, respectively.    
\emph{Lower:} sketch of our model of the southern arc (not to scale). 
The synchrotron radio-to-optical radiation is located within 
the compact MERLIN emitter, whereas  IC X-ray emission is produced
in a more extended region.
\label{sketch}}
\end{figure}

The FR~II galaxy 4C74.26 is located at redshift $z = 0.104$ 
($\sim$0.5~Gpc from Earth)\footnote{Throughout this paper we use the 
cosmology $H_0 = 71$~km~s$^{-1}$~Mpc$^{-1}$, $\Omega_0 = 1$ and 
$\Lambda_0 =  0.73$. One arcsecond represents$1.887$~kpc on the plane of 
the sky at $z = 0.104$.}.  Two X-ray sources separated by
$\sim$10$^{\prime\prime}$  were detected with  Chandra
\cite[by][]{Erlund_TwoShocks} at the termination region of the
southern jet, as shown in Figure~\ref{sketch} (upper).  In the present
work we study the southern  X-ray source, with a luminosity
$L_{\rm x} \sim 10^{41}$~erg~s$^{-1}$ at 2~keV and called  ``the southern arc''. 
The shape of this emission is arc-like with a characteristic size 
$l_{\rm x} \sim 10^{\prime\prime}$, and encloses a compact radio source.

\subsection{The southern arc}

Compact radio emission from the southern arc was detected 
with the MERLIN high resolution interferometer ($\nu_{\rm r} = 1.66$~GHz) 
with a flux $f_{\rm r} \sim$0.04~Jy, and luminosity 
$L_{\rm r} \sim$ 1.9$\times$10$^{40}$~erg~s$^{-1}$ per unit logarithmic 
bandwidth $\delta \nu = \nu$. This emission is located in a   
region of width $l_{\rm r} < 1^{\prime\prime}$ on the plane of the sky.
In addition, faint and diffuse radiation was 
detected at IR ($\nu_{\rm ir} =$ 1.36$\times$10$^{14}$~Hz) and optical 
($\nu_{\rm opt} =$ 6.3$\times$10$^{14}$~Hz) bands, with fluxes 
$\sim$8.4$\times10^{-6}$ and  2.82$\times$10$^{-7}$~Jy,
respectively, and located in a region of width $\gtrsim l_{\rm r}$. 
However, there is a linear structure (in both bands) that traces the 
brightest edge
of the MERLIN radio emission, and seems to be cupped within it.

Two factors indicate that the  southern arc of X-ray emission is not 
synchrotron.
First, $l_{\rm x} > l_{\rm r}$ is inconsistent with the X-ray emitting electrons 
being more energetic and therefore cooling more rapidly
as they advect away from the shock.
Second, the steep spectrum between IR and optical 
\citep[see Fig.~13 in][]{Erlund_TwoShocks} indicates the maximum
energy of (synchrotron) emitting electrons. We note that
similar characteristics are observed in other sources 
\cite[e.g.][]{sources}.

\cite{Erlund_TwoShocks}
suggested that the multi-wavelength emission from the southern arc 
is produced by non-thermal electrons, emitting synchrotron 
radiation from radio to optical, and up-scattering the cosmic microwave 
background (CMB) photons (with energy $\sim$7$\times10^{-4}$~eV
and energy density $U_{\rm cmb} = $6$\times$10$^{-13}$~erg~cm$^{-3}$) 
to the X-ray domain. 
In this scenario,  the radio-to-IR spectral index is $\alpha = 0.75$
(the synchrotron flux density at frequency $\nu$ is 
$f_{\nu} \propto \nu^{-\alpha}$), which corresponds to $p = 2.5$
in the non-thermal electron energy distribution when 
the emission is produced in the same volume.

\section{The hotspot as a magnetic field damping region}

In this work we consider the same emission mechanisms as in 
\cite{Erlund_TwoShocks}, but allow the synchrotron and Inverse
Compton (IC) emission  to be 
produced in regions with different spatial extents. 
In particular; electrons accelerated at the shock emit
synchrotron radiation from  radio to optical in a compact region 
behind the shock,  whereas the 
IC X-ray emission is located in an extended region.
In Fig.~\ref{sketch} (lower) we sketch our model.

The synchrotron (s) and IC cooling length of electrons with Lorentz factor 
$\gamma$  is $l_{\rm s,ic}(\gamma) = t_{\rm s,ic}(\gamma) v_{\rm sh}/r$, where 
$t_{\rm s,ic}(\gamma)$ is the cooling timescale. 
The shock velocity is approximately the same as the jet velocity which we 
take characteristically to be $v_{\rm sh} = 10^{10}$~cm~s$^{-1}$ 
($\sim c/3$ and Lorentz factor $\Gamma_{\rm sh} \sim 1.06$) 
in line with observations 
of similar objects \cite[see][and references therein]{Steenbrugge_08}.
 We use $r = 7$ as the shock compression ratio 
for a  non-relativistic shock whose downstream thermal pressure is 
dominated by relativistic electrons, although $r \sim 4$ may 
still apply if non-relativistic ions dominate the pressure downstream of the
shock \citep{Kirk_00}. Our conclusions
are not sensitive to the exact value of $r$.

\subsection{Inverse Compton  X-ray emission}

The IC X-ray emission is produced by electrons with 
$\gamma_{\rm x} \sim 10^3$ and 
$l_{\rm ic}(\gamma_{\rm x})\sim10^{4} (v_{\rm sh}/10^{10}\,{\rm cm\,s^{-1}})$~arcsec,
which is much larger than $l_{\rm x}$.  
The synchrotron cooling length $l_{\rm s}(\gamma_{\rm x})$
is also greater than $l_{\rm x}$, unless the magnetic 
field in the X-ray emitting region is $\sim 360$~$\mu$G. 
However, such a large magnetic field would produce synchrotron radio 
radio emission much brighter than $L_{\rm r}$ (see next section).
Furthermore, we show below that the amplified magnetic field, of the order
of $\sim$100~$\mu$G, is confined to a small volume close to the shock.
Therefore, adiabatic expansion is probably the dominant cooling mechanism 
as the particles flow out of the hotspot.

Unless $\Gamma_{\rm sh} \gtrsim 10$, X-ray emitting electrons are 
non-thermal and follow a power law energy distribution 
$\propto \gamma^{-p}$ \cite[e.g.][]{Mixed_Ne}.   
Assuming that
the X-ray emitting volume is $V_{\rm x} \sim 300$~arcsec$^{3}$, 
the energy density of these non-thermal electrons is
$\sim$10$^{-9} (\gamma_{\rm min}/50)^{-0.5}$~erg~cm$^{-3}$
where the power law terminates at a minimum Lorentz factor $\gamma_{\rm min}$.
The magnetic field with the same energy density  is
$\sim 100(\gamma_{\rm min}/50)^{-0.25}$~$\mu$G. 
These results correspond to the case where $p = 2.5$
(see however Sect.~\ref{ir}). In the following sections we take
$100$~$\mu$G as a fiducial magnetic field since it represents equipartition 
between magnetic and relativistic electron energy densities and is typical
of other hotspots \cite[e.g.][]{B_equip}.

\subsection{Synchrotron  emission}

Considering that $\gamma(\nu) \sim 4.5\times10^{-4}(\nu/B)^{0.5}$
is the Lorentz factor of electrons emitting synchrotron radiation at 
$\nu$ in a magnetic field $B$, $l_{\rm s}$ can be written as
\begin{equation}
\frac{l_{\rm s}(\nu)}{[\prime\prime]} \sim 12
\left(\frac{\nu}{\rm GHz}\right)^{-0.5} 
\left(\frac{B}{100\,{\rm \mu G}}\right)^{-1.5}
\left(\frac{v_{\rm sh}}{10^{10}\,{\rm cm\,s^{-1}}}\right).
\label{l_cool}
\end{equation}

\subsubsection{Radio}
\label{radio}

MERLIN emitting electrons have 
$\gamma_{\rm r} \equiv \gamma(\nu_{\rm r})\sim 
2\gamma_{\rm x}(B/100\,\mu{\rm G})^{-0.5}$. 
If both radio and X-ray emission are produced by non-thermal
electrons that follow the same power-law energy distribution, 
$L_{\rm x}/L_{\rm r} \sim (\gamma_{\rm x}/\gamma_{\rm r})^{3-p}(U_{\rm cmb}/U_{\rm mag})
V_{\rm x}/V_{\rm r}$  and
$B \sim 100$~$\mu$G corresponds to $V_{\rm x}/V_{\rm r} \sim 5\times10^3$, 
where $U_{\rm mag} = B^2/8\pi$ and $V_{\rm r}$ is the volume of the  
synchrotron emitter\footnote{Note
that if $V_{\rm x} = V_{\rm r}$, an unrealistically small magnetic field of 
$0.3$~$\mu$G would be needed to explain the observed fluxes.}. 
Such a large ratio between emitting volumes is not implausible provided the 
magnetic field is
inhomogeneous in the shock downstream region and the  synchrotron 
emitter consists of features smaller than the MERLIN point spread function
(FWHM $0.15^{\prime\prime}$) as seen in parts of the MERLIN data. 

The synchrotron cooling length  of MERLIN emitting electrons is
$l_{\rm s}(\nu_{\rm r}) \sim 9.3^{\prime\prime}(B/100\,{\rm \mu G})^{-3/2}
(v_{\rm sh}/10^{10}\,{\rm cm\,s^{-1}})$,
and a very large magnetic field of
$\sim 2.4 \,(v_{\rm sh}/10^{10}\,{\rm cm\,s^{-1}})^{2/3}$~mG 
would be required to  match 
$l_{\rm s}(\nu_{\rm r}) = 0.1^{\prime\prime} \sim l_{\rm r}$.
This result suggests that the downstream extent of the compact emission 
detected at $\nu_{\rm r}$ is not the result of fast synchrotron cooling, 
as we can confirm when we take into account the IR emission.

\subsubsection{Infrared}
\label{ir}

The synchrotron cooling length  of IR emitting electrons is
$l_{\rm s}(\nu_{\rm ir}) \sim 0.03^{\prime\prime}(B/100\,{\rm \mu G})^{-3/2}
(v_{\rm sh}/10^{10}\,{\rm cm\,s^{-1}})$, indicating that these particles 
radiate most of their energy within $l_{\rm r}$.
(This angular distance is not resolved by the IR observations.)
This is consistent with a radio-to-IR electron  energy spectral index of 
$p \sim 2.5$  with the cooling break in the spectrum occuring 
close to IR wavelengths\footnote{Note that the relationship 
$p = (r + 2)/(r -1)$ breaks down for
mildly relativistic shocks \citep{Kirk_00, Bell_11} or when non-linear 
feedback is important \cite[e.g.][]{Amato_05}.}. 
Note that if the emitting volume were determined by synchrotron 
cooling, $l_{\rm s}(\nu) \propto \nu^{-0.5}$ giving $p = 2\alpha = 1.5$ 
since $\alpha$ is measured to be $0.75$. 
This very hard spectrum is unlikely since it diverges toward high energy
and would be remarkable in hotspots,
supporting the conclusion that the downstream radio extent $l_{\rm r}$
must be determined by factors other than synchrotron cooling.
As we discuss in Sect.~\ref{mfa} this may be the result of the damping of 
the magnetic field \cite[see][for a review]{Klara_review}.

\subsubsection{Optical}

Optical emission produced by synchrotron radiation of 
electrons with $\gamma(\nu_{\rm opt}) \sim \gamma(\nu_{\rm ir})$
is almost co-spatial with the IR emission, and this explains the 
linear structure cupped within $l_{\rm r}$\footnote{The faint diffuse 
IR and optical emission may be the result of
CMB photons up-scattered by electrons with $\gamma \sim 50$.}.
The synchrotron turnover $\nu_{\rm c}$ between $\nu_{\rm ir}$ and $\nu_{\rm opt}$ 
indicates that the maximum energy of non-thermal electrons is
$E_{e,\rm max} \sim \gamma(\nu_{\rm c}) m_ec^2 \sim 
0.3 (\nu_{\rm c}/\nu_{\rm ir})^{0.5} (B/100\mu{\rm G})^{-0.5}$~TeV.

\section{Magnetic field amplification}
\label{mfa}

The amplification of the magnetic field at strong shocks in supernova
remnants was demonstrated by \cite{Vink-Laming} and \cite{Berezhko-03},
deriving the magnetic field from $l_{\rm s}$ downstream of the shock.
A theoretical explanation was provided by \cite{Bell_04} 
showing that non-resonant hybrid instabilities are capable of enhancing
the magnetic field by orders of magnitude. 
Magnetic field amplification is also responsible for 
$B \sim 100$~$\mu$G  in the southern arc in 4C74.26 since   
it is much larger than the expected value in the jet upstream of the 
termination shock \cite[e.g.][]{B_jet}.

Bohm diffusion (electron mean free path $\lambda \sim r_{\rm g}$) in a 
$\sim$100~$\mu$G magnetic field would be expected to 
accelerate electrons with synchrotron X-ray emitting energies as seen in 
supernova remnants \citep{Stage_06}.  
However, $\nu_{\rm c} \sim \nu_{\rm ir,opt}$
determined by a competition between shock acceleration and synchrotron cooling
indicates that
acceleration is slow and therefore that the electron diffusion coefficient $D$
is much larger than the Bohm value $D_{\rm Bohm}$:  
\begin{equation}
\frac{D}{D_{\rm Bohm}} \sim 10^6 \left(\frac{v_{\rm sh}}{10^{10}\,{\rm cm\,s^{-1}}}
\right)^2
\left(\frac{\nu_{\rm ir}}{\nu_{\rm c}}\right), 
\end{equation}
independent of $B$ \cite[see e.g.][]{Casse}.
Such a large diffusion coefficient in an amplified magnetic field
is expected if it is structured on a scale $s$ much smaller than the
Larmor radius of the electrons being accelerated.
Small angle scattering by magnetic field randomly orientated in cells 
of size $s$ produces $D\sim (r_{\rm g}/s) D_{\rm Bohm}$
and then
\begin{equation}
\frac{s}{\rm cm} \sim 10^{7} \left(\frac{\nu_{\rm c}}{\nu_{\rm ir}}\right)^{1.5}
\left(\frac{B}{100\,{\rm \mu G}}\right)^{-1.5}
\left(\frac{v_{\rm sh}}{10^{10}\,{\rm cm\,s^{-1}}}\right)^{-2}.
\label{lambda}
\end{equation}
In comparison the ion skin-depth is 
$c/\omega_{\rm pi} \sim 2.3 \times10^9 \,(n/10^{-4}\,{\rm cm^{-3}})^{-0.5}$~cm,
where $n$ is the particle density downstream of the shock (assumed to be 
$7$ times the  jet density),   and 
\begin{eqnarray}
\frac{s}{c/\omega_{\rm pi}} \sim & 0.01 
\left(\frac{\nu_{\rm c}}{\nu_{\rm ir}}\right)^{1.5}
\left(\frac{v_{\rm sh}}{10^{10}\,{\rm cm\,s^{-1}}}\right)^{-2}   \nonumber\\
{}& \left(\frac{B}{100\,{\rm \mu G}}\right)^{-1.5}
\left(\frac{n}{10^{-4}\,{\rm cm^{-3}}}\right)^{0.5}.
\end{eqnarray}

Given the uncertainties in the parameter values, the approximate nature of 
the theoretical models, and the wide range of the spatial scales ($s$, 
$r_{\rm g}$, $l_{\rm r}$), it is not significant or surprising that our estimate 
of $s/(c/\omega_{\rm pi})$ differs from unity by a factor of $\sim 0.01$. 
The order of magnitude similarity of $s$ 
and $c/\omega_{\rm pi}$ supports the contention that 
shock-generated small-scale turbulence scatters non-thermal electrons during
diffusive shock acceleration. This is consistent with \cite{Sironi_11}  
who discuss the various processes related to the Weibel instability
that excite turbulence on the characteristic scale of $c/\omega_{\rm pi}$.
Simulations show that magnetic field generated by the Weibel instability 
decays downstream of the shock because of its relatively small scalelength
\citep{Sironi_11,Bret_13, Sironi_13}.
This would account for the cut-off of synchrotron emission in 4C74.26 
far short of the synchrotron cooling distance of radio-emitting electrons
($l_{\rm r} \ll l_{\rm s}(\nu_{\rm r})$).
These electrons continue up-scattering CMB photons, thus producing IC
X-ray emission downstream of the shock after the MERLIN
radio emission has ceased.

\subsection{Limit on ions maximum energy}

Since the response of highly relativistic ions is similar to that 
of electrons with the same energy in a tangled amplified magnetic field,
we can expect protons to have a similar ratio of $D/D_{\rm Bohm}$.
Protons can be accelerated to higher energies than electrons because their 
radiative losses are minimal,
but the maximum energy to which they are accelerated is limited
because their acceleration time is increased by the ratio $D/D_{\rm Bohm}$.  
The Hillas parameter $v_{\rm sh}\,B\,R$ \citep{Hillas}, where 
$R \sim 2^{\prime\prime}$ ($\sim 2.8$~kpc) is the 
characteristic length of the source, would suggest proton acceleration to 
$\sim$100~EeV in the termination shock of 4C74.26,
but the maximum energy is reduced  to only $\sim$100~TeV
if $D\sim 10^6\,D_{\rm Bohm}$ since the Hillas parameter  assumes 
$D\sim D_{\rm Bohm}$ and is otherwise reduced by the factor $(D/D_{\rm Bohm})^{-1}$.
Another perspective on the same effect is that the mean free path for
scattering by small-scale turbulence $\lambda \sim r_{\rm g}^2/s$ is larger 
than the size of the system if $s \sim c/\omega_{\rm pi}$ and $r_{\rm g}$
is the Larmor radius of an EeV proton.  
This result suggests that the mildly relativistic termination shock in
4C74.26 is a poor accelerator of UHECR.

\section{Conclusions}

We model the radio to X-ray emission  in the 
southern hotspot of the FR~II galaxy 4C74.26. 
Our study is based on three key observational features:
1) Compact MERLIN emission region: it is too thin to be the result of
fast synchrotron cooling (Sect.~\ref{radio}). 
2) The radio to IR spectrum ($\alpha = 0.75$) is too flat for the emitting
volume to be determined 
by synchrotron cooling through this wavelength range (Sect.~\ref{ir}).
3) The turnover of the synchrotron spectrum at IR/optical frequencies
requires $\lambda \gg r_{\rm g}$ for any reasonable shock velocity
(Sect.~\ref{mfa}).
These three features fit well  in a scenario in which the MERLIN radio 
emission traces out the region where the magnetic field is amplified
by plasma instabilities with small length scale (e.g. Weibel).

The magnetic field 
in equipartition with non-thermal electrons in the MERLIN  emission region 
is  $\sim$100~$\mu$G and similar to the values obtained by other authors 
\cite[e.g.][]{B_equip}. An unrealistically large magnetic field
$\sim 2.4\,(v_{\rm sh}/10^{10}\,{\rm cm\,s^{-1}})^{2/3}$~mG would be needed to
explain the compact radio emission in terms of synchrotron cooling.  
If $B \sim 100$~$\mu$G  in the synchrotron  emission region,
the maximum energy of non-thermal electrons is $\sim$0.3~TeV.

If ions are accelerated 
as well, protons with energy $\sim$0.3~TeV diffuse also with mean free path
$\lambda \gg r_{\rm g}$. If $\lambda$ is similarly larger than the Larmor 
radius at higher proton energies,
then the maximum proton energy at the termination shock of 4C74.26
is only $100$~TeV instead of the $100$~EeV indicated by the Hillas parameter.
This may have important implications for the understanding of the origins 
of UHECR.

\acknowledgments
We thank the referees for the constructive reports.  
The research leading to this article has received funding
from the European Research Council under the European
Community's Seventh Framework Programme (FP7/2007-2013)/ERC grant agreement 
no. 247039. We acknowledge support from the UK Science and Technology 
Facilities Council under grant number ST/K00106X/1.

\end{document}